\documentclass{aa}
\input psfig.sty
\usepackage[english]{babel}
\usepackage[latin1]{inputenc}
\usepackage{graphicx}
\usepackage{psfrag}
\usepackage{subfigure}
\usepackage{rotating}
\usepackage{lscape}
\usepackage{latexsym}
\usepackage{array}
\usepackage{times}
\usepackage{german}

\begin{document}
   \title{Dynamical Evolution of High Velocity Clouds in the Intergalactic Medium}

   \subtitle{}

   \author{C.~Konz \inst{1,2}, C.~Br\"uns \inst{3,4} \and G.T.~Birk \inst{1,2}
          }

   \offprints{C.~Konz}

   \institute{Institut f\"ur Astronomie und Astrophysik der Universit\"at
M\"unchen, Scheinerstra\3e 1, D-81679 M\"unchen, Germany
 \and 
  Center of Interdisciplinary Plasma Science, Garching, Germany
 \and
Radioastronomisches Institut der Universit\"at Bonn, Auf dem H\"ugel 71, D-53121
 Bonn, Germany
 \and
The Australia Telescope National Facility, CSIRO, PO Box 76, Epping NSW 1710,
Australia
             }

   \date{Received x x, 2002; accepted }
\mail{konz@usm.uni-muenchen.de}

   \abstract{
 \ion{H}{i} observations of high-velocity clouds (HVCs) indicate, that they are
 interacting with their ambient medium. Even clouds located in the very outer
 Galactic halo or the intergalactic space seem to interact with their ambient
 medium. In this paper, we investigate the dynamical evolution of high velocity 
 neutral gas clouds moving through a hot magnetized ambient plasma by means of 
 two-dimensional magnetohydrodynamic plasma-neutral gas simulations. 
 This situation is representative for the fast moving dense neutral gas 
 cloudlets in the Magellanic Stream as well as for high velocity clouds in 
 general. The question on the dynamical and thermal stabilization of a cold 
 dense neutral cloud in a hot thin ambient halo plasma is numerically investigated. The 
 simulations show the formation of a comet-like head-tail structure combined 
 with a magnetic barrier of increased field strength which exerts a stabilizing 
 pressure on the cloud and hinders hot plasma from diffusing into the cloud. 
 The simulations can explain both the survival times in the intergalactic medium
 and the existence of head-tail high velocity clouds.
   }

   \maketitle
%

\section{Introduction}

High-velocity clouds (HVCs) -- first discovered by Muller et al. (\cite{muller}) 
-- are defined as neutral atomic hydrogen clouds with radial velocities that 
cannot be explained by simple galactic rotation models.
After almost 40 years of eager investigations there is still no general 
consensus on the origin and the basic physical parameters of HVCs
(e.g.~Bregman \cite{bregman}).
The most critical issue of HVC research is the distance uncertainty. 
Danly et al. (\cite{danly}), Keenan et al. (\cite{keenan}) and Ryans
et al. (\cite{ryans}) consistently determined an upper distance limit of 
d $\leq$  5 kpc to HVC complex M.
The most important step forward is the very recently determined distance bracket
of 4 $\leq$ d $\leq$ 10 kpc towards HVC complex A by van Woerden et al. 
(\cite{van Woerden}). These results clearly place the HVC complexes M 
and A in the gaseous halo of the Milky Way.

Parallel to the growing evidence that a significant fraction of the HVC 
complexes are located in the Milky Way halo, Blitz et al. (\cite{blitz}) 
supported the hypothesis that some HVCs are of extragalactic origin.
They argued, that it is reasonable to assume that primordial gas -- left
over from the formation of the Local Group galaxies -- may appear as HVCs.
Braun \& Burton (\cite{bb99}) identified 65 compact and isolated HVCs
and argued that this ensemble represents a homogeneous subsample of HVCs at
extragalactic distances.
Observational evidence for extragalactic HVCs may also be found by the 
detection of the highly ionized high-velocity gas clouds by Sembach et al. 
(\cite{sembach}), because of its very low pressure of about p k$^{-1} \approx 
5 \ \mathrm{K cm}^{-3}$. 

The Magellanic Stream (MS) and the Leading Arm (LA) (Putman et al.
\cite{putman}) both form coherent structures over several tens of degrees 
having radial velocities in the HVC regime. They represent debris 
most likely caused by the tidal interaction of the Magellanic Clouds with the 
Milky Way. The distances to these features are of the order of 50 kpc.
 
Meyerdierks (\cite{meyerdierks}) detected a HVC that appears like a cometary 
shaped cloud with a central core and an asymmetric envelope of warm neutral atomic 
hydrogen (the particular HVC is denoted in literature as HVC A2). He
interpreted this head-tail structure as the result of an interaction between 
the HVC and normal galactic gas at lower velocities.
Towards the HVC complex C, Pietz et al. (\cite{pietz96}) discovered the so-called 
{\sc Hi} ``velocity bridges'' which seem to connect the HVCs with the normal 
rotating interstellar medium.
The most straight forward interpretation for the existence of such
structures is to assume that a fraction of the HVC gas was 
stripped off the main condensation.
Br\"uns et al. (\cite{bruens00}) extended the investigations of Meyerdierks 
(\cite{meyerdierks}) and Pietz et al. (\cite{pietz96}) over the entire sky 
covered by the new Leiden/Dwingeloo {\sc Hi} 21-cm line survey (Hartmann \& Burton 
\cite{hartmann97}) and found head-tail structures in all HVC-complexes including
the Magellanic Stream, except 
for the very faint HVC-complex L. Their analysis
revealed that the absolute value of the radial velocity of the tail is always 
lower than the value for the head of the
HVC ($|v_{\rm LSR,tail}| < |v_{\rm LSR,head}|$). In addition, it was shown that 
the fraction of HVCs showing a head-tail structure increases proportional to 
the peak column density and increasing radial velocity $|v_{\rm GSR}|$.

HVCs mostly appear as 
``pure'' neutral atomic hydrogen clouds. Absorption line studies provide 
information on the ionization state and the metalicity of HVCs. The results 
indicate that the bright and very extended HVC complexes consist 
(at least partly) of processed material, having $\le$ 1/3 of the solar 
abundances (Wakker \cite{wakker-met}). \\
Thus, the question of the dynamical and thermal stabilization of a cold dense
neutral cloud in a hot thin ambient plasma arises. To confine a HVC in the
Magellanic Stream by pressure a relatively high halo gas density is required
(Mirabel et al.~\cite{mirabel}, Weiner \& Williams \cite{weiner}). 
Therefore, the lifetime of the HVCs should be significantly limited by
evaporation (Murali \cite{murali})
since the necessary pressure for confinement is associated
with a high energy transfer.
An alternative mechanism of dynamical stabilization is the magnetic confinement
of the cloud. In this contribution, we numerically investigate the formation of
a magnetic barrier around the HVC in the Magellanic Stream and its stabilizing 
effects on the neutral gas
cloud. 

\section{Observations}

The gaseous arms of the Magellanic System are very interesting objects to study
the evolution of clouds moving with high velocities relative to an ambient 
medium. The Magellanic Stream (MS) is a coherent structure which starts from 
the Magellanic Clouds and trails over 100$^{\circ}$ passing the southern 
Galactic pole. 
The physical parameters of both the gas in the Magellanic Stream and the ambient
medium must support survival times of the order 10$^9$ yr 
predicted by numerical simulations (Gardiner \cite{gardiner}).

\begin{figure}[]
  \centerline{
\psfig{figure=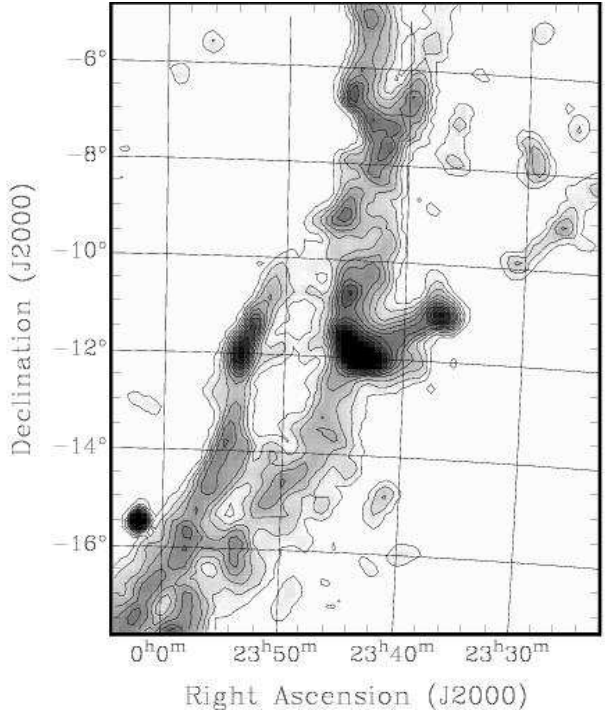,width=8.8cm,angle=0}}
  \caption[]{This figure shows the column density distribution of a part of the
  Magellanic Stream observed with the multi-beam facility of the Parkes 
  telescope. The contour levels represent 0.1, 1, 2, 3, ...$\cdot10^{19} 
  {\rm cm}^{-2}$.}
 \label{magteil}
\end{figure}

The low distance of about 50 kpc enables studies of the distribution of gas on
scales of about 200 pc with single dish telescopes.
Recently, the Parkes multi-beam narrow-band survey of the entire Magellanic
System was completed (see Br\"uns et al.~\cite{bruens00b}).
Fig. \ref{magteil} shows exemplary the column density distribution of a part of the 
MS, also known as MS~IV. The Stream is subdivided in two filaments
containing a number of condensations. Some of them show a horseshoe like
structure. In addition, the Stream is accompanied by a number of isolated clouds
at similar velocities that are most likely part of the gaseous feature.
The detailed analysis of the gas within the MS is difficult, as some clouds are
superposed on the same line of sight. The analysis of the isolated clouds is
much easier in this respect. Observable quantities are the angular extent, the
line-widths, and the column density distribution. Other important values like the
gaseous mass, the density etc. can only be calculated, if the distance to the
clouds is known. We assume a distance of 50 kpc to all clouds in this section. 

Typical line-widths derived from isolated clouds are $FWHM \approx$ 25 -- 30 
km~s$^{-1}$, typical diameters are of the order 40'. This corresponds to 580 pc
at a distance of 50 kpc. A particle moving with a velocity of 
$v \approx$ 15~km~s$^{-1} \approx$ 15 kpc Gyr$^{-1}$ crosses the cloud in
$\approx$ 0.04 Gyr, i.e. the crossing time is much smaller than the age of the
Magellanic Stream. The virial mass of a cloud can be calculated using 
$M_{\rm vir}$ = 210~r~($\Delta$v)$^2$, where r is the radius of the cloud in pc
and $\Delta$v is the linewidth in km~s$^{-1}$. Virial masses for clouds in the
MS are $M_{\rm vir}$ = 3.94 10$^7$ (r~/~300pc) ($\Delta$v~/~25km~s$^{-1}$)$^2$ 
M$_{\sun}$, while observed H{\small I} masses are of the order few $\times$
10$^4$ M$_{\sun}$. Apparently, the observed H{\small I} masses are
orders of magnitude to low to provide gravitational stabilization.
Consequently, the cloud must be confined by external pressure. 
Direct observations of the ambient medium provide only weak constraints on the
physical parameters. The density of the ambient medium is of the order 
$n_{\rm IGM} \approx$ 10$^{-5}$ cm$^{-3}$ (Murali \cite{murali}; Weiner \& Williams 
\cite{weiner}; Mirabel et al.~\cite{mirabel}). 
The temperature of the gas in the outer Galactic halo is
expected to be similar to the temperature in the lower Galactic halo ($T_{\rm IGM}
\approx$ 1 -- 2 $\times$ 10$^6$ K). For a fully ionized medium, the pressure
can be calculated using $P_{\rm IGM}$~k$^{-1}$~=~2.3~$n_{\rm IGM}$~$T_{\rm IGM}$, 
i.e. $P_{\rm IGM}$~k$^{-1}$~$\approx$ 50~K~cm$^{-3}$. 

The pressure within the MS can be estimated using the correlation between 
linewidth, turbulent velocity and kinetic Temperature

\parbox{\columnwidth}{
\begin{equation} \label{eq:linewidth}
\Delta v = 2 \left( ln 2 \left( \frac{2 k T_k}{m_H} + V^2 \right) \right)^{1/2}
\end{equation}
}

\noindent where $\Delta v$ is the linewidth, $T_k$ is the kinetic temperature, k is the
Boltzmann constant, m$_H$ is the mass of the hydrogen atom, and $V$ is the most
probable turbulent velocity. The assumption that there is no turbulence 
provides a possibility to derive an upper limit to 
the kinetic temperature (a.k.a. Doppler temperature) from Eq.~
(\ref{eq:linewidth}): $T_D$ = 21.8~($\Delta v$)$^2~\approx$~2$\times$10$^4$~K, 
where $T_D$ is in K and $\Delta v$ is $FWHM$ in km~s$^{-1}$. The actual
kinetic temperature must be lower, as processes like turbulence are likely to 
occur. Typical observed volume densities within the Magellanic Stream 
are of the order $n$ = few $\times$ 10$^{-2}$~cm$^{-3}$, assuming a distance of 
50 kpc for the Magellanic Stream. 
Using $P k^{-1}  = n T $ allows to estimate the pressure in the
Stream to be of the order $P k^{-1} \approx $ few $\times 100 \ 
\mathrm{K cm}^{-3}$.

Pressure equilibrium between the neutral gas in the clouds and the
ionized medium is possible, as both pressures are only rough estimates. 
In addition, the contribution of magnetic fields to the question of stability is
completely unknown.

There are a number of clouds in the Magellanic Stream and the Leading Arm
showing up with head-tail structures, i.e. asymmetries in the column density
distribution associated with velocity gradients (Br\"uns et al. \cite{bruens00}). 
A detailed analysis of these features is extremely complicated, because of
the superposition of clouds along the same line of sight.

\begin{figure}[]
  \centerline{
\psfig{figure=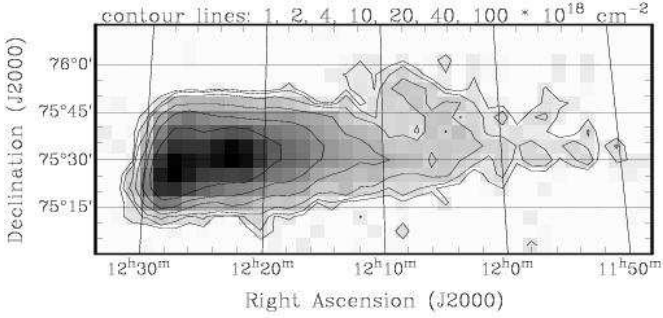,width=8.8cm,angle=0}}
\caption[]{This figure shows the column density distribution of HVC125+41-207.
  Contour lines represent 1, 2, 4, 10, 20, 40, 100 $\times$10$^{18}$~cm$^{-2}$.
  This clouds shows a very prominent head-tail structure.}
 \label{hvc125}
\end{figure}

Another interesting class of HVCs are the so-called compact, isolated HVCs.
Braun \& Burton (\cite{bb99}) argue, that these HVCs are located within the
intergalactic medium of the Local Group. They are by definition compact {\em
and} isolated -- a detailed analysis of the H{\small I} distribution is
therefore much easier compared with the Magellanic Stream.
Fig. \ref{hvc125} shows the column density distribution of a compact and 
isolated high-velocity cloud with a distinct head-tail structure. 
This cloud was discussed in detail in Br\"uns et al.~ 
(\cite{bruens01}). The distance to this cloud was determined to be of the
order 130 kpc, i.e. the cloud is located in the very outer halo of the Milky Way 
or the near intergalactic space. We consider this cloud as a generic example for
the class of head-tail HVCs investigated in this contribution.

\begin{figure}[]
\centerline{
\psfig{figure=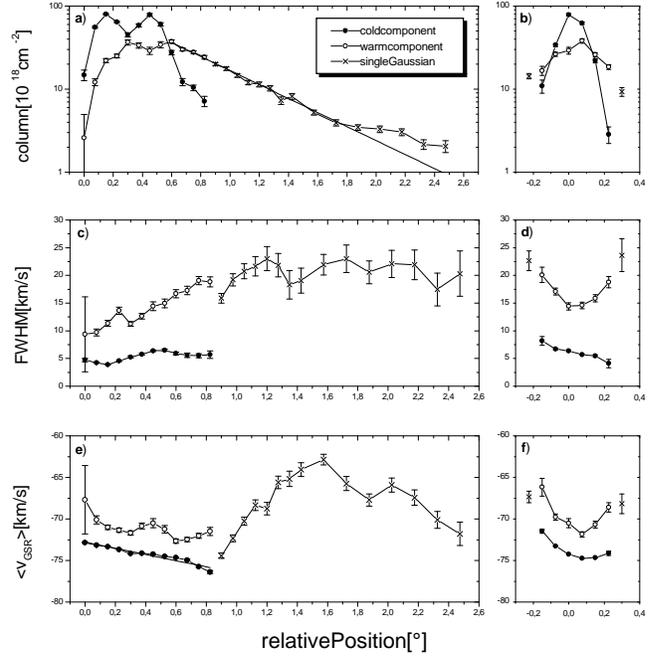,width=8.8cm,angle=0,bbllx=30pt,bblly=90pt,bburx=540pt,bbury=595pt}}
  \caption[]{This figure shows the result from a Gaussian decomposition of the
  H{\small I} data of HVC125+41-207 (Br\"uns et al. \cite{bruens01}). The cold 
  gas-phase is represented by filled 
  circles, the warm gas-phase by open circles. Results from a single Gaussian 
  fit are represented by crosses. The diagrams {\bf a},{\bf c} and {\bf e} show 
  the results for a slice along the tail, the diagrams {\bf b},{\bf d} and 
  {\bf f} represent the slice perpendicular to the orientation of the tail. 
  The line in diagram {\bf a} is an exponential fit to the data.
  The line in diagram {\bf e} is a fit to the velocity field.}
 \label{gauss125}
\end{figure}

Fig. \ref{gauss125} shows the result from a Gaussian decomposition of the
H{\small I} data of HVC125+41--207. 
About 12.5 \% of the H{\small I}-gas is located in the tail. The spectra show 
in general two components, one with a low and one with a high velocity 
dispersion (i.e. a cold and a warm gas-phase). The velocity dispersion 
of the warm H{\small I}-gas is increasing from FWHM $\approx$~10 $\rm km~s^{-1}$~ 
to FWHM $\approx$~20 $\rm km~s^{-1}$~ over the extend of the cloud and remains 
approximately constant in the tail. This corresponds to a Doppler 
temperature increasing
from $T_{\rm D}$ = 2000 K to $T_{\rm D}$ = 10000 K in the tail. 
There is also a trend of higher velocity dispersions towards the outskirts 
of the cloud. In addition, the radial velocities $|v_{\rm GSR}|$ of the warm 
component are lower than the velocities in the colder gas. This is a strong
indication that the warm gas is in the process of being stripped off the main
body of the HVC.
The observed morphology of HVC125+41--207 clearly demonstrates, that this cloud
is currently interacting with its ambient medium similar to the HVCs in the
Magellanic Stream and Leading Arm. 

Br\"uns (in prep.) performed deep H{\small I} observations of a sample of 10 
compact HVCs with the 100-m Effelsberg telescope. {\em All} of them show up with
a head-tail structure -- some of the tails have a fairly low column density 
($\approx$10$^{18}$~cm$^{-2}$). 

Altogether, the observations demonstrate, that HVCs in the Galactic halo 
or in the near intergalactic space interact with their ambient medium producing
head-tail structures.

\section{Numerical simulations of high velocity clouds}

\begin{figure}[htbp]
\psfrag{HVC}[][]{\textbf{\Huge{HVC}}}
\psfrag{v}[][]{\huge{v}}
\psfrag{magnetic}[][]{\LARGE{magnetic}}
\psfrag{field lines}[][]{\LARGE{field lines}}
\psfrag{ionosphere}[][]{\LARGE{ionosphere}}
\psfrag{x}[][]{\huge{x}}
\psfrag{y}[][]{\huge{y}}
\psfrag{z}[][]{\huge{z}}
\includegraphics[angle=0,width=88mm,keepaspectratio]{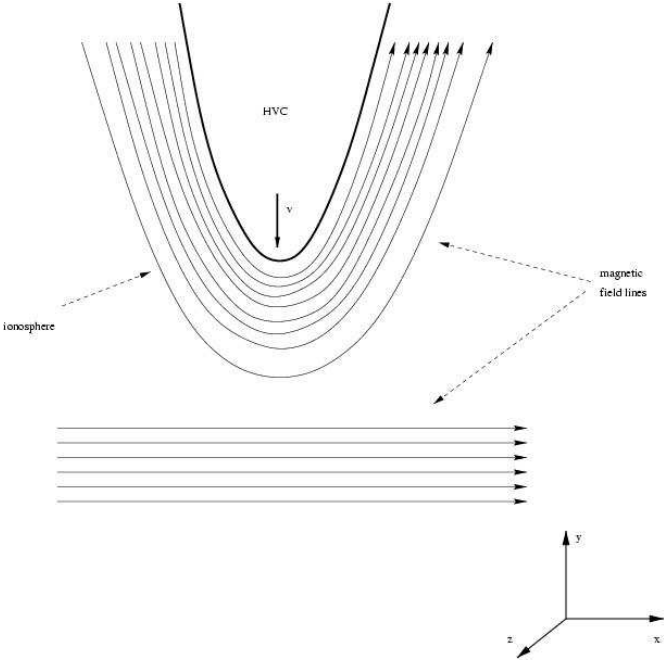}
\caption{A high velocity cloud moving through a magnetized plasma} 
\label{fig:hvc}
\end{figure}

\begin{figure}[htbp]
\includegraphics[angle=0,width=88mm,keepaspectratio]{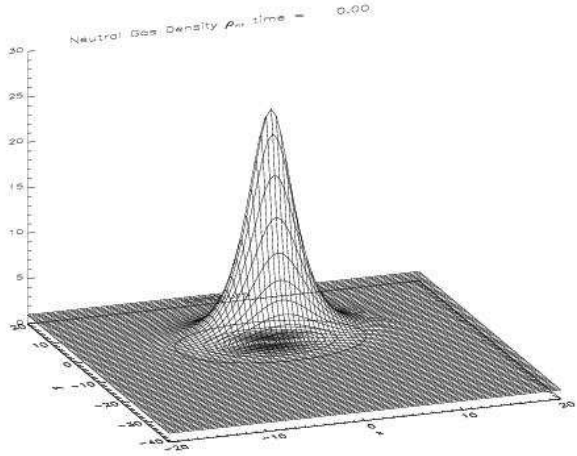}
\caption{The initial neutral gas density \( \rho_{\mathrm{n}} \); figures are in
non-dimensional units} 
\label{fig:dn0}
\end{figure}

\begin{figure}[htbp]
\includegraphics[angle=0,width=88mm,keepaspectratio]{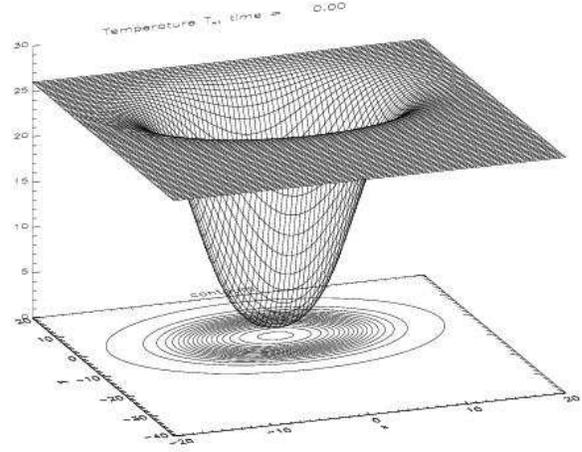}
\caption{The initial temperature profile \( T_{\mathrm{n}} \) for the neutral gas} 
\label{fig:tn0}
\end{figure}

\begin{figure}[htbp]
\includegraphics[angle=0,width=88mm,keepaspectratio]{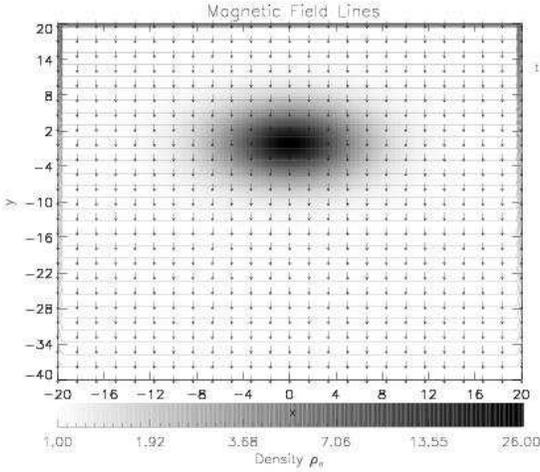}
\caption{The initial configuration for the neutral gas density, the plasma flow,
and the magnetic field} 
\label{fig:mag0}
\end{figure}

\begin{figure}[htbp]
\includegraphics[angle=0,width=88mm,keepaspectratio]{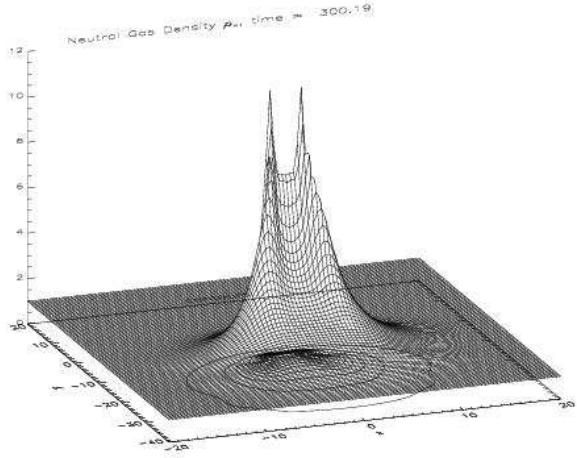}
\caption{The neutral gas density at \(t \approx 300 \tau_{\mathrm{A}} \) for the
non-magnetic case} 
\label{fig:dn50}
\end{figure}

\begin{figure}[htbp]
\includegraphics[angle=0,width=88mm,keepaspectratio]{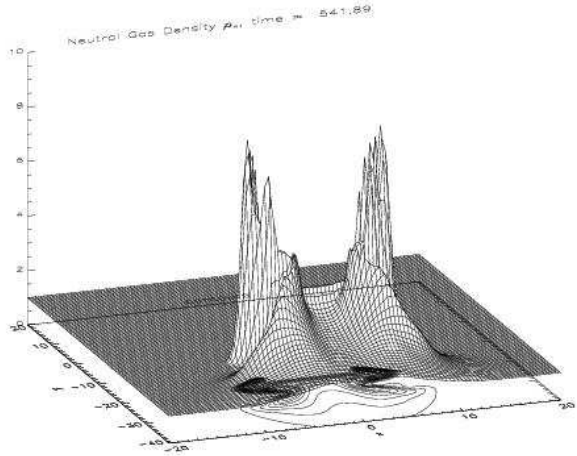}
\caption{The neutral gas density at \(t \approx 542 \tau_{\mathrm{A}} \) for the
non-magnetic case} 
\label{fig:dn90}
\end{figure}

\begin{figure}[htbp]
\includegraphics[angle=0,width=88mm,keepaspectratio]{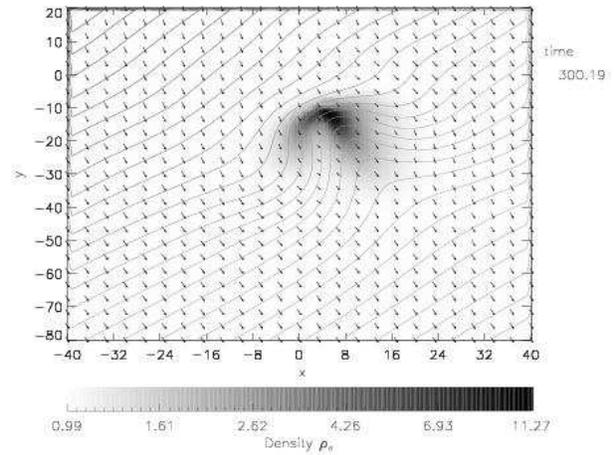}
\caption{The neutral gas density, the plasma flow, and the magnetic field  at
\(t \approx 300 \tau_{\mathrm{A}} \) for the case of an inclined inflow magnetic
field} 
\label{fig:magrhonlog50}
\end{figure}

\begin{figure}[htbp]
\includegraphics[angle=0,width=88mm,keepaspectratio]{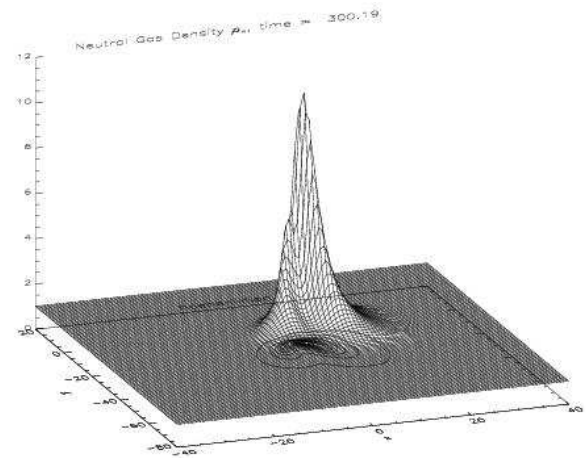}
\caption{The neutral gas density at \(t \approx 300 \tau_{\mathrm{A}} \) for an
inclined inflow magnetic field} 
\label{fig:rhon50}
\end{figure}

\begin{figure}[htbp]
\includegraphics[angle=0,width=88mm,keepaspectratio]{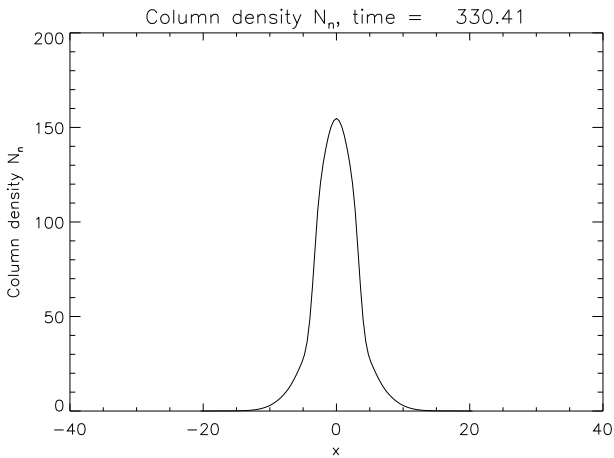}
\caption{A neutral gas column density cut along the \(x\)-axis at \(t \approx
330 \tau_{\mathrm{A}} \)} 
\label{fig:nn55}
\end{figure}

In order to study the dynamical evolution of a fast moving, cool, dense neutral
gas cloud in a magnetized, hot ambient medium we performed 2-D MHD
plasma-neutral gas simulations assuming invariance in the \(z\)-direction
(\(\partial{}/ \partial{z} = 0\)). \\
In the simulations, a dense neutral gas cloud is moving in the \(y\)-direction
impinging on a homogeneous magnetic field lying in the \(x\)-direction (see
Fig.~\ref{fig:hvc}). \\
Here, we consider, for example, physical parameters for HVCs in the Magellanic
Stream, i.e. the outer Galactic halo.  
For the high velocity clouds MSI-VI in the Magellanic Stream the particle
density of the ambient plasma lies in the range of \(10^{-5}
\mbox{cm}^{-3}\) -- \( 7 \times 10^{-4} \mbox{cm}^{-3}\)
(Weiner \& Williams \cite{weiner}; Mirabel et al.~\cite{mirabel})
for the Galactic halo and is even lower -- at about \( 10^{-6}
\mbox{cm}^{-3}\) -- in the intergalactic medium. 
The temporal evolution of the plasma-neutral gas system is calculated by
numerically integrating the following set of normalized fluid equations (for a
derivation of the fluid equations see Birk \& Otto (\cite{birk})): \\
\parbox{\columnwidth}{
\begin{equation} \label{eq:pcontinuity}
\frac{\partial{\rho}}{\partial{t}} = - \vec{\nabla} \cdot \left( \rho \vec{v}
\right)
\end{equation}
\makebox[\columnwidth][r]{\textit{(plasma continuity equation)}}
}
\smallskip
\parbox{\columnwidth}{
\begin{equation} \label{eq:ncontinuity}
\frac{\partial{\rho_{\mathrm{n}}}}{\partial{t}} = - \vec{\nabla} \cdot \left(
\rho_{\mathrm{n}} \vec{v}_{\mathrm{n}} \right)
\end{equation}
\makebox[\columnwidth][r]{\textit{(neutral gas continuity equation)}}
}
\smallskip
\parbox{\columnwidth}{
\begin{eqnarray} \label{eq:pmomentum}
\frac{\partial{}}{\partial{t}} \left( \rho \vec{v} \right) & = & - \vec{\nabla}
\cdot \left( \rho \vec{v} \vec{v} \right) - \frac{1}{2} \vec{\nabla} p + \left(
\vec{\nabla} \times \vec{B} \right) \times \vec{B} \nonumber \\
& &  - \rho 
\nu^{\mathrm{S}}_{12} \left( \vec{v} - \vec{v_{\mathrm{n}}} \right)
\end{eqnarray}
\makebox[\columnwidth][r]{\textit{(plasma momentum equation)}}
}
\smallskip
\parbox{\columnwidth}{
\begin{equation} \label{eq:nmomentum}
\frac{\partial{}}{\partial{t}} \left( \rho_{\mathrm{n}} \vec{v}_{\mathrm{n}}
\right)  =  - \vec{\nabla} \cdot \left( \rho_{\mathrm{n}} \vec{v}_{\mathrm{n}}
\vec{v}_{\mathrm{n}} \right) - \frac{1}{2} \vec{\nabla} p_{\mathrm{n}} -
\rho_{\mathrm{n}} \nu^{\mathrm{S}}_{21}
\left( \vec{v}_{\mathrm{n}} - \vec{v} \right) 
\end{equation}
\makebox[\columnwidth][r]{\textit{(neutral gas momentum equation)}}
}
\smallskip
\parbox{\columnwidth}{
\begin{equation} \label{eq:induction}
\frac{\partial{\vec{B}}}{\partial{t}}  =  \vec{\nabla} \times \left( \vec{v}
\times \vec{B} \right)
\end{equation}
\makebox[\columnwidth][r]{\textit{(induction equation)}}
}
\smallskip
\parbox{\columnwidth}{
\begin{equation} \label{eq:ppressure}
\frac{\partial{p}}{\partial{t}}   =  - \vec{v} \cdot \vec{\nabla} p - \gamma p
\vec{\nabla} \cdot \vec{v} 
 - 3 \left( \gamma - 1 \right) \nu^{\mathrm{S}}_{12} \left( p - \frac{\rho}{
\rho_{\mathrm{n}} } p_{\mathrm{n}} \right)
\end{equation}
\makebox[\columnwidth][r]{\textit{(plasma pressure equation)}}
}
\smallskip
\parbox{\columnwidth}{
\begin{eqnarray} \label{eq:npressure}
\frac{\partial{p_{\mathrm{n}}}}{\partial{t}} & = & - \vec{v}_{\mathrm{n}} \cdot
\vec{\nabla} p_{\mathrm{n}} - \gamma_{\mathrm{n}} p_{\mathrm{n}} \vec{\nabla}
\cdot \vec{v}_{\mathrm{n}} \nonumber \\
& & - 3 \left( \gamma_{\mathrm{n}} - 1 \right)
\nu^{\mathrm{S}}_{21}
\left( p_{\mathrm{n}} - \frac{\rho_{\mathrm{n}}}{ \rho} p \right)  \; .
\end{eqnarray}
\makebox[\columnwidth][r]{\textit{(neutral gas pressure equation)}}
}
The quantities \( \rho \), \( p \), \( \vec{v} \), \( \rho_{\mathrm{n}} \), \(
p_{\mathrm{n}} \), \( \vec{v}_{\mathrm{n}} \), \( \vec{B} \), \( e \), \( \gamma
\), and
\( \gamma_{\mathrm{n}} \) denote the plasma mass density, pressure, and
velocity, the neutral gas mass density, pressure, and velocity, the magnetic
field, the elementary charge, and the ratio of specific heats for the plasma and
the neutral gas, respectively. The term \( \vec{v} \vec{v} \) represents the
dyadic product \( \left( v_{i} v_{j} \right)_{i,j} \). The ratio of specific
heats has been chosen as \( \gamma = \gamma_{\mathrm{n}} = 5/3 \). \\
In the presented simulations, we assume a
continuous ionization equilibrium. So far, we
are not interested in ionization phenomena, but in the dynamical evolution of
the high velocity cloud, only. \\
The frequencies \( \nu^{\mathrm{S}}_{12} \) and \( \nu^{\mathrm{S}}_{21} \)
denote effective elastic collision frequencies. In
order to guarantee momentum conservation, they have to fulfil the following
relation (for details see Birk \& Otto (\cite{birk})): \\
\begin{equation} \label{eq:nu12s}
\nu^{\mathrm{S}}_{12} \rho = \nu^{\mathrm{S}}_{21} \rho_{\mathrm{n}}
\end{equation}
The symmetric collision
frequency~\( \nu^{\mathrm{S}}_{12} \) is a measure for the relative collisional
friction between the plasma and the neutral gas. Heuristically, we chose the
reasonable ansatz
\begin{equation} \label{eq:nuscal}
\nu^{\mathrm{S}}_{12} = \nu^{\mathrm{S}}_{0} \rho_{\mathrm{n}} T^{1/2}
\end{equation}
for the collision frequency~\( \nu^{\mathrm{S}}_{12} \) where \( T = p / \rho \)
is the normalized plasma temperature and \( \nu^{\mathrm{S}}_{0} \) is a
constant factor. \\
All quantities \( X \) are normalized to typical values \( \hat{X} \) of the
system. Here we chose parameters typical for high velocity clouds in the
Magellanic Stream and neutral gas clouds in the intergalactic medium.
Length scales are normalized to a typical extension of a cloud of \( \hat{L} = 100
\; \mbox{pc} \approx 3 \times 10^{20} \; \mbox{cm} \). The particle density
(ions and neutral atoms, respectively) is normalized to a value of \( \hat{n} =
10^{-3} \; \mbox{cm}^{-3} \).  The magnetic field is normalized to \(
\hat{B} = 3 \; \mu \mbox{G} \). The actual magnetic field in the Galactic halo
is not known. Upper limits from cosmic ray confinement, pulsars, etc. indicate
field strengths of some microGauss (Vall\'ee \cite{vallee}, Beck et 
al.~\cite{beck}). We perform the
simulation with an initial field strength of one percent of the normalization
field. Our findings show that even with this relatively weak initial field a
substantial dynamical stabilization of the HVC results.
We assume a monoatomic neutral gas,
e.g. hydrogen atoms, such that the ion mass and the neutral atom mass are
equal. Furthermore, we assume only one species of single ionized ions.\\
One may note that a pure hydrodynamic treatment is
barely applicable to HVCs in the Galactic halo since the fluid
approximation does not strictly hold. \\ 
With the above choice for \( \hat{L} \), \( \hat{n} \), and \( \hat{B} \) further
normalizations follow in a generic way. The mass densities~\( \rho \) and \(
\rho_{\mathrm{n}} \) are normalized to \( \hat{\rho} = \left( m_{\mathrm{e}} +
m_{\mathrm{i}} \right) \hat{n} \approx m_{\mathrm{i}} \hat{n} = m_{\mathrm{a}} \hat{n}
\) where \( m_{\mathrm{e}} \), \( m_{\mathrm{i}} \), and \( m_{\mathrm{a}} \) are
the electron, ion, and neutral atom mass, respectively. Velocities are
normalized to the Alfv\'en velocity \( v_{\mathrm{A}} = \hat{B} / \sqrt{4 \pi
\hat{\rho} } \), frequencies to the inverse Alfv\'en time \( 1/ \tau_{\mathrm{A}}
= v_{\mathrm{A}} / \hat{L} \), and pressures to \( \hat{p} = \hat{B}^{2} / 8 \pi
\). From the normalized equation of state for ideal gases \( p = n T \) the
temperature of reference can be deduced as \( \hat{T} = \hat{B}^{2} / 16 \pi \hat{n}
k_{\mathrm{B}} \) where \( k_{\mathrm{B}} \) is the Boltzmann constant. \\
For the above values of the system we find a mass density \( \hat{\rho} \approx
1.67 \times 10^{-27} \; \mbox{g} \, \mbox{cm}^{-3} \), the Alfv\'en velocity~\(
v_{\mathrm{A}} \approx 210 \; \mbox{km} \, \mbox{s}^{-1} \), the Alfv\'en
transit time~\( \tau_{\mathrm{A}} \approx 1.31 \times 10^{13} \; \mbox{s}
\approx 416 \, 000 \; \mbox{yr} \), and the temperature~\( \hat{T} \approx 1.3 \times
10^{6} \; \mbox{K} \). \\
For numerical reasons, the density gradient of the modeled HVC is chosen
to be smaller than a realistic one. Consequently, the neutral gas temperature in
the neutral cloud is unrealistically high. However, the dynamics are not
significantly influenced by this choice since the mechanical forces are due
rather  to pressure than to density or temperature gradients. On the whole, we
model the neutral gas as if its ionization energy was high enough to keep it 
neutral at \( 10^{6}  \mbox{K} \). The ionization energy does not enter the 
calculation at any point, therefore being an arbitrary choice. \\
The integration of the plasma-neutral gas equations is done on an equidistant
2-D grid by a second order leapfrog scheme where the partial derivatives are
realized as finite differences by the FTCS method (Forward Time Centered
Space). The continuity, momentum and pressure equations are integrated with a
flux correction of the source terms between the two half time steps while the
induction equation is integrated by means of a semi-implicit algorithm
(Dufort-Frankel). Details of the numerical code can be found in Birk \& Otto
(\cite{birk}). \\
A schematic drawing of the simulation is given in Fig.~\ref{fig:hvc}. In order
to simulate the infall of a compact dense neutral gas cloud perpendicular to a
magnetic field in a hot plasma we first transform to the neutral gas rest
frame. Then the initial configuration is defined by a dense neutral gas cloud
centered at the origin \( \left( x,y \right) = \left( 0,0 \right) \) of the system having
the form 
\begin{equation} \label{eq:cloud}
\rho_{\mathrm{n}} \left( r \right) = \rho_{\mathrm{min}} + \frac{
\rho_{\mathrm{n}_{0}} }{ \cosh{a r}} \; .
\end{equation}
Here, \( r \) gives the distance from the origin of the system while \( a = 1/2  \)
defines the scale length for the neutral gas density gradient. The constants \(
\rho_{\mathrm{min}} \) and \( \rho_{\mathrm{n}_{0}} \) define the minimum and
the maximum neutral gas density and are chosen to be \( \rho_{\mathrm{min}} = 1
\) and \( \rho_{\mathrm{n}_{0}} = 25 \), respectively. The initial neutral gas
density is shown in Fig.~\ref{fig:dn0}.
The neutral gas cloud is initially at rest, i.e., \( \vec{v_{\mathrm{n}}} = 0
\). The neutral gas is assumed to be initially in pressure equilibrium at a
homogeneous pressure \( p_{\mathrm{n}_{0}} = T_{\mathrm{n}_{0}} \left(
\rho_{\mathrm{min}} + \rho_{\mathrm{n}_{0}} \right) \) where \(
T_{\mathrm{n}_{0}} = 1 \) is the minimum temperature of the neutral gas
cloud. The temperature profile is then given by \(
T_{\mathrm{n}} \left( r \right) = p_{\mathrm{n}_{0}} / \rho_{\mathrm{n}} \left(
r \right) \)(see Fig.~\ref{fig:tn0}).
The neutral gas outside the cloud is about a factor 25 hotter than at the center
of the cloud. This temperature gradient, however, is still small compared to
realistic temperature gradients by factors of \( 10^{3}\) -- \( 10^{4} \) in
typical HVCs. This is due to numerical reasons in so far as stronger gradients
require an enhanced spatial and temporal resolution which makes a simulation
extremely time- and memory-consuming. However, simulations with steeper
gradients in the temperature and density of the neutral gas component have shown
qualitatively the same evolution of the system apart from the longer calculation
time. Since we are interested in the longterm evolution of a plasma-neutral gas
system we restrain ourselves to intermediate temperature and
density gradients. \\
The plasma itself initially has a homogeneous density~\( \rho_{0} = 1 \) and a
similar temperature profile as the neutral gas:
\begin{equation} \label{eq:temp}
T \left( r \right) = T_{0} - \frac{ T_{0} - T_{\mathrm{n}_{0}} }{ \cosh{a r}} \;
.
\end{equation}
Here, \( T_{0} = 10 \) is the maximum temperature of the plasma. It is
worthwhile mentioning that the external neutral gas is even hotter than the external
plasma in this simulation. The neutral gas outside the cloud just serves as a
reservoir of heat for the outer plasma which is heated up in the course of time
via the collisional \( \nu_{12}^{\mathrm{S}} \)-term in the pressure
equation. It is not dense enough to influence the motion of the plasma
significantly due to the small value of the collision frequency~\(
\nu^{\mathrm{S}}_{0} = 8 \times 10^{-3} \). During the dynamical evolution, the
neutral
gas gathers the same velocity profile as the plasma motion. Initially, the
plasma streams in the \(y\)-direction with the velocity profile
\begin{equation} \label{eq:vel}
v_{y} \left( r \right) = v_{y_{0}} \left( 1 - \frac{1}{ \cosh{ a r }} \right)
\end{equation}
where \( v_{y0} = -0.1 \) is the minimum velocity of the plasma.
This leaves the center of the neutral cloud at rest while plasma is streaming
past the neutral cloud in the negative \(y\)-direction and across an initially
homogeneous magnetic field in the \(x\)-direction: \( B_{x} = B_{x0} = 0.01 \),
\( B_{y} = B_{z} = 0 \). Since the local Alfv\'en velocity is given by \(
v_{\mathrm{A}} = B / \sqrt{ 4 \pi \rho} \) this plasma flow is superalfv\'enic
at the beginning of the simulation due to the small magnetic field. This means
that the plasma flow is able to compress and deform the magnetic field
significantly. However, there is no shock to be found because of the high
temperature \( T \approx 1.3 \times 10^{7} \; \mbox{K} \) and, connected with
it, the high velocity of sound~\( c_{\mathrm{s}}=\sqrt{ 2 \gamma k_{\mathrm{B}}
T / m_{\mathrm{i}} } \approx 661 \; \mbox{km} \, \mbox{s}^{-1} \). \\
The complete initial configuration is shown in Fig.~\ref{fig:mag0}.
The neutral gas density is shown in logarithmic scaling while the lines indicate
magnetic field lines and the arrows indicate the plasma flow. \\
At the boundaries of the simulation box all quantities are extrapolated to the
first order of the Taylor expansion. There are three exceptions, though. The
plasma and neutral gas densities and the collision frequency~\(
\nu_{12}^{\mathrm{S}} \) are chosen to be symmetric at both
\(y\)-boundaries. Furthermore, the magnetic field component~\( B_{x} \) is
chosen to be symmetric at the \(y_{\mathrm{max}}\)-boundary because of numerical
stability reasons. The most important exception, however, is given by the fact,
that the plasma flow is maintained at the upper \(y\)-boundary by setting \(
v_{y} \left( y_{\mathrm{max}} \right) = v_{y0} \). Altogether, we get a
plasma-neutral gas system where magnetic flux, plasma, neutral gas, and energy
can freely cross the \(x\)-boundaries but where magnetic flux and mass flux are
partly fixed at the \(y\)-boundaries. \\
The simulations presented here are all ideal in the sense that no resistivity apart
from a small constant background resistivity \( \eta_{0} = 10^{-5} \) is applied
due to numerical reasons. \\
The calculations are done in a 2-D box with \(x\) going from \(-20\) to \(20\)
and \(y\) going from \(-40\) to \(20\). We use a uniform grid with \(103\) grid
points in the \(x\)- and \(153\) grid points in the \(y\)-direction resulting in
a constant grid spacing of \(0.4\). \\
\begin{figure}[htbp]
\includegraphics[angle=0,width=88mm,keepaspectratio]{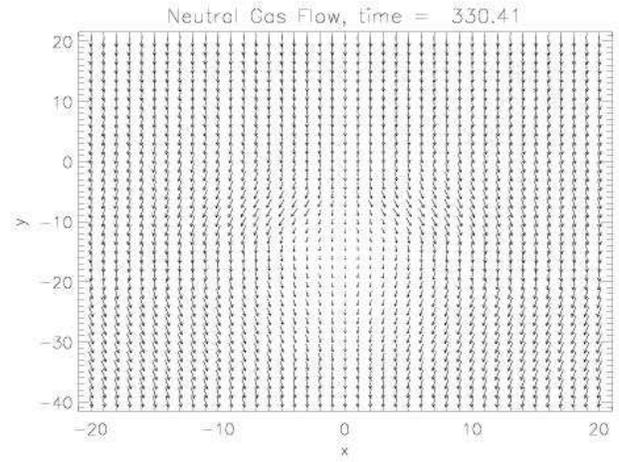}
\caption{The neutral gas flow at \( t \approx 330 \tau_{\mathrm{A}} \)} 
\label{fig:flow55}
\end{figure}
Since the system is treated ideally in a magnetohydrodynamic view the magnetic
field lines are frozen into the plasma. Thus, they are deformed in the course of
time by the plasma streaming past the neutral gas cloud. At the beginning of the
simulation, both the neutral gas and the plasma at the origin are at rest. The
collisional terms \( \rho \nu_{12}^{\mathrm{S}} \left( \vec{v} -
\vec{v_{\mathrm{n}}} \right) \) and \( \rho_{\mathrm{n}} \nu_{21}^{\mathrm{S}}
\left( \vec{v_{\mathrm{n}}} - \vec{v} \right) \) in the momentum equations lead
to a deceleration of the plasma impinging upon the neutral gas cloud and to an
acceleration of the neutral gas cloud itself. However, the acceleration of the
neutral gas cloud is rather small due to the high neutral gas density of the
cloud. Outside the cloud, plasma and neutral gas densities are comparable such
that the neutral gas acquires the same velocity as the streaming plasma. \\
The neutral gas flow is shown in Fig.~\ref{fig:flow55}. One finds a velocity
pattern that resembles the initial velocity pattern apart from the fact that the
center of the neutral gas cloud has moved towards the lower
boundary. Furthermore, the plasma and neutral gas is now flowing around the
cloud as if it was a soft obstacle in a hydrodynamic flow. The center of
the neutral gas cloud is now moving with a small velocity in the negative
\(y\)-direction as a result of the continuous impact of plasma on the cloud. The
exchange of momentum between the incoming plasma and the resting neutral gas
cloud is performed via elastic ion-neutral gas collisions. The momentum exchange
leads to an acceleration of the cloud in the initial rest frame, i.e., to a
deceleration of the cloud in the plasma rest frame. This is consistent with
ultraviolet absorption observations hinting at a deceleration of the HVCs as
they approach the Galactic disk (Benjamin \cite{benj}; Danly \cite{dan}). \\
\begin{figure}[htbp]
\includegraphics[angle=0,width=88mm,keepaspectratio]{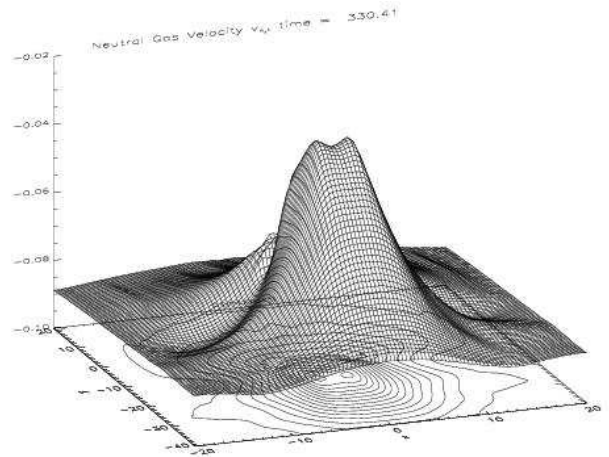}
\caption{The neutral gas velocity~\( v_{\mathrm{n}_{y}} \) at \( t \approx 330
\tau_{\mathrm{A}} \)} 
\label{fig:vny55}
\end{figure}
Fig.~\ref{fig:vny55} shows the neutral gas velocity~\( v_{\mathrm{n}_{y}} \) at
the same time as the flow in Fig.~\ref{fig:flow55}. As already
mentioned the center of the cloud is now moving with a small but finite velocity
in the negative \(y\)-direction. \\
The simulation time of about \( 330 \) Alfv\'en
times corresponds to a time interval of about \( 140 \) million years which is
comparable to crossing times of HVCs in the Magellanic Stream. \\
\begin{figure}[htbp]
\includegraphics[angle=0,width=88mm,keepaspectratio]{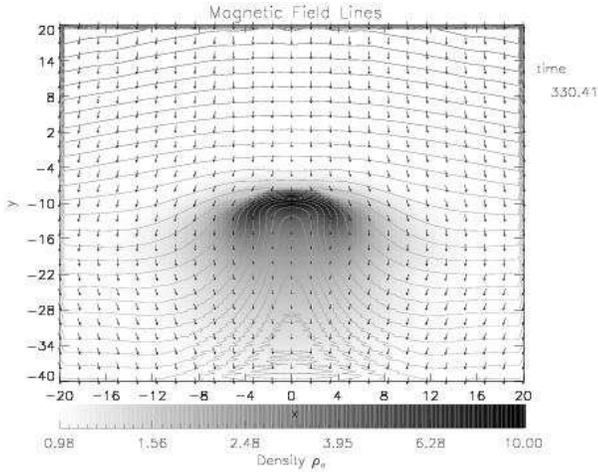}
\caption{The neutral gas density, the plasma flow, and the magnetic field  at \(
t \approx 330 \tau_{\mathrm{A}} \)} 
\label{fig:mag55}
\end{figure}
A complete representation of the system after about \( 330 \) Alfv\'en times is
given in Fig.~\ref{fig:mag55}. The arrows represent the flow of the plasma
around the neutral gas cloud. Since the cloud does not act as a solid but as a
soft obstacle there are no real stagnation points in front and behind the
cloud. Nevertheless, the plasma flow encountering the cloud is partly redirected
sideways similar to a hydrodynamic stream around a solid obstacle. Since the
magnetic field lines are frozen in the plasma, magnetic flux is transported via
the plasma flow towards the front edge of the cloud while the field lines on both
sides of the cloud are draped around the cloud forming a magnetic barrier at the
front and a magnetotail at the wake of the HVC. \\ 
\begin{figure}[htbp]
\includegraphics[angle=0,width=88mm,keepaspectratio]{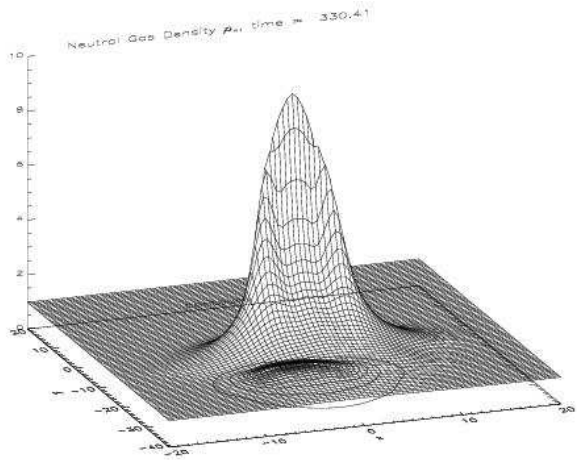}
\caption{The neutral gas density at \(t \approx 330 \tau_{\mathrm{A}} \)} 
\label{fig:dn55}
\end{figure}
The formation of a magnetic barrier around the head of the neutral gas cloud
resembles very much the formation of a magnetic barrier at the front of a HVC
via the critical velocity effect (Konz et al. \cite{konz}). In turn, the
formation of a magnetotail is quite similar to the formation of a magnetotail by
the well-studied interaction of the solar wind with unmagnetized planets
(Luhmann \cite{luhma}; Luhmann et al. \cite{luhmb}) and comets (Formisano et
al. \cite{form}; McComas et al. \cite{comas}; Flammer et al. \cite{flam}). \\
The neutral gas cloud itself is slightly deformed into a U-like structure,
visible in Fig.~\ref{fig:mag55} in logarithmic scaling of the neutral gas
density, together with a weak tail consisting of neutral gas being slowly stripped
off the wings of the cloud. The stripping of neutral gas off the edges of the
cloud results from the momentum transfer from the halo gas to the neutral gas
particles. The stability of the cloud is not influenced significantly by this
ongoing effect during the time period under consideration. Eventually, this
abrasion of neutral gas forms a tail. The phenomenon of tail formation is
consistent with the observation (see Fig.~\ref{gauss125}). The upper column
of this figure shows that the tail mainly consists of the warm component while
the colder core of the neutral gas cloud remains unaffected by the stripping.
After the formation of the magnetic barrier the resulting density profile of the 
cloud, as it
is shown in Fig.~\ref{fig:dn55}, remains almost constant in time as later
time-steps of the simulation demonstrate. The HVC almost keeps its width
and its maximum density. \\
The dynamical stability of the neutral gas cloud is a direct effect of the
magnetic confinement by the accumulated magnetic barrier. This can be shown by
performing the same simulation without any magnetic field. Figs.~\ref{fig:dn50}
and \ref{fig:dn90} show the neutral gas density at about \( 300
\tau_{\mathrm{A}} \) and \( 540 \tau_{\mathrm{A}} \) for this case.
One finds the initially compact neutral gas cloud being first disrupted into
two major parts and then completely falling apart. The cold neutral cloud is
unstable against impact of hot plasma. The momentum and energy transfer from the
incoming plasma to the front of the cloud leads to a disruption of the cloud
into several parts. \\
For comparison to the case of a magnetic field initially perpendicular to the
flow direction, Figs.~\ref{fig:magrhonlog50} and \ref{fig:rhon50} show the cloud
and the magnetic field configuration for the case of an inclined magnetic
field with initial field strengths \( B_{x0} = B_{y0} = 0.03 \). As in the
previous simulation, the initial plasma inflow is purely in the negative
y-direction.
As long as there is a significantly strong component of the magnetic field
perpendicular to the flow direction the formation of a magnetic
barrier stabilizes the cloud in two ways. First of all, the increased magnetic
field in the barrier exerts a magnetic pressure~\( \vec{B}^{2}/ 8 \pi \) on the
cloud and partly compensates the ram pressure ~\( \rho \vec{v}^{2} /2 \) of the
plasma stream. Since the system is ideal in the magnetohydrodynamic description
plasma cannot move across magnetic field lines. Therefore, the plasma flow is
redirected by the magnetic barrier to stream around the HVC. On the inner side of
the magnetic barrier, the magnetic pressure is compensated by the internal
plasma pressure.\\ 
As expected, we find a lower limit of the magnetic field strength sufficient to
stabilize the HVC. For the parameters chosen, this limit is about \( 15 \mathrm{nG}
\) meaning one order of magnitude higher than the mean intergalactic magnetic
field (Vall\'ee \cite{valleeb}). 
At the beginning of the simulation, the plasma density was set
to be homogeneous. However, there is a pressure gradient due to the fact that
the plasma inside the cloud region is much cooler than outside the cloud. Due to
the finite temperature of a HVC and neutral gas ionization via UV photons,
collisions and soft X-ray background there is always a small fraction of plasma
inside a HVC amounting to the order of several percent of the total density. In
the course of the simulation, this fraction is increased up to \( 50 \% \) by
plasma streaming towards the center of the cloud along the field lines, i.e.,
along the tail and by a compression of the whole cloud due to the external
plasma pressure. The plasma inside the cloud, however, is cooled by energy
transfer to the cold neutral gas and would be subject to recombination events if
recombination was included in the simulation. So far, recombination is not
considered in our simulations. However, recombination of the cold plasma inside
the cloud would further increase the build-up of a magnetic barrier because the
plasma pressure gradient would be even steeper due to the loss of plasma. That
means that more magnetic flux is transported inside the magnetic barrier and the
magnetic tail by the plasma heading for the cold center of the cloud. The
magnetic pressure finally stops this plasma flow and isolates the plasma inside
the cloud from the ambient halo plasma. In our simulation, the plasma inside the
cloud takes up to \( 50 \% \) of the total density. This percentage, however,
only reaches such a high level because of the relatively small density of the
neutral gas cloud compared to the density of the ambient plasma. For a higher
neutral gas density the plasma fraction inside the cloud only amounts to a few
percent. \\
Altogether, we have a system consisting of three different regions: an outer
region of hot ambient plasma exerting the ram pressure on the region of the
magnetic barrier and the inner region with cold plasma and neutral gas
compensating the magnetic pressure from inside. Elastic collisions inside the
cloud are much more frequent than outside since the higher neutral gas density
overcompensates the lower temperature in Eq.~(\ref{eq:nuscal}). Therefore, plasma
and neutral gas inside the cloud have the same temperature and the same
velocity. \\
\begin{figure}[htbp]
\includegraphics[angle=0,width=88mm,keepaspectratio]{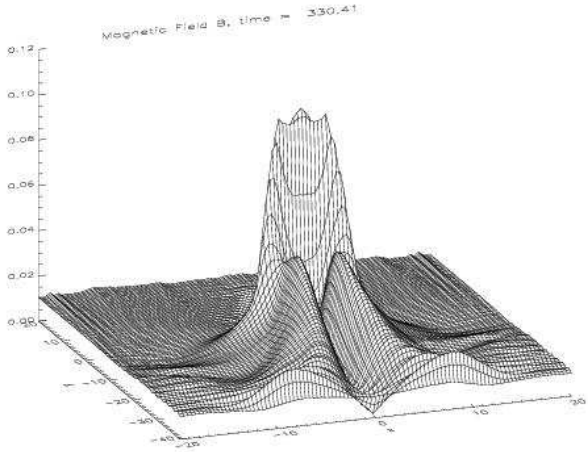}
\caption{The magnetic field strength at \(t \approx 330 \tau_{\mathrm{A}} \)} 
\label{fig:b55}
\end{figure}
The magnetic barrier not only dynamically stabilizes the HVC but also
thermally. The thermal conduction perpendicular to a magnetic field is
drastically reduced compared to the thermal conduction parallel to the magnetic
field (Braginskii \cite{brag}). The coefficient~\(  \kappa_{\perp} \) for the
heat conduction perpendicular to the magnetic field decreases with increasing
magnetic field strength as \( B^{-2} \) (Van der Linden \& Goossens
\cite{van}). Fig.~\ref{fig:b55} shows the magnetic field strength after about \(
330 \tau_{\mathrm{A}} \). At the front of the HVC, the magnetic field strength
is increased by a factor \( 10 \), in the tail by a factor \( 4 \). This means
that the perpendicular heat conduction is reduced by a factor \( 1/100 \) or \(
1/16 \), respectively. In our simulation, heat conduction was not included so
far. Since the main part of the temperature gradient is
perpendicular to the magnetic field lines no qualitative changes
are to be expected if heat conduction is included. \\
\begin{figure}[htbp]
\includegraphics[angle=0,width=88mm,keepaspectratio]{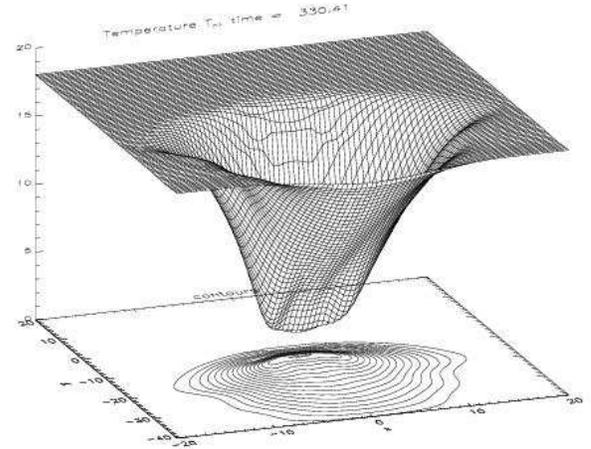}
\caption{The neutral gas temperature at \(t \approx 330 \tau_{\mathrm{A}} \)} 
\label{fig:tn55}
\end{figure}
Heat conduction is suppressed and plasma cannot cross the magnetic barrier
the temperature of the neutral gas cloud remains almost constant in
time. Fig.~\ref{fig:tn55} shows the temperature profile of the neutral gas after
about \( 330 \tau_{\mathrm{A}} \). Although a little bit deformed the initial
temperature profile is almost maintained. Steep temperature gradients are found
in the region of the magnetic barrier surrounding the minimum temperature at the
maximum of the neutral gas density. The temperature on the outside is still by a
factor \( 10 \) higher than on the inside of the cloud. The magnetic barrier
therefore insulates the cloud against heat exchange and heat conduction. \\
In the tail, the temperature gradient is much smaller leading to a tail at an
intermediate temperature. This relatively warm neutral gas in the tail shows a
higher velocity dispersion than the core of the cloud itself, a feature that has
been confirmed by measuring the velocity dispersion both in the head and in the
tail of HVCs (see Sect. 2). \\
\begin{figure}[htbp]
\includegraphics[angle=0,width=88mm,keepaspectratio]{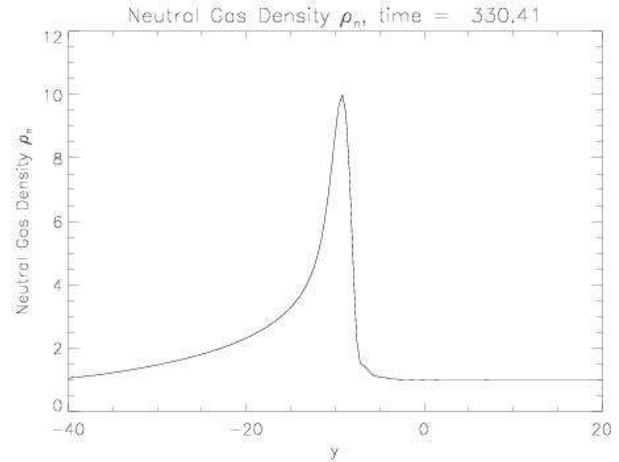}
\caption{A neutral gas density cut along the \(y\)-axis at \(t \approx 330
\tau_{\mathrm{A}} \)} 
\label{fig:dncut55}
\end{figure}
How is the neutral gas density in the simulation compared to
observations of head-tail structures? Fig.~\ref{fig:dncut55} shows a cut along
the \(y\)-axis for the neutral gas density, i.e., along the axis of
symmetry. In the plasma rest frame, the neutral gas cloud is moving to the
right, i.e.,
in the positive \(y\)-direction. The density cut shows a steep gradient at the
head of the cloud where the cloud is well confined by the magnetic barrier and a
smooth almost exponential decay towards the tail. In this presentation, the HVC
looks like a comet
trailing in its wake a tail of material that is stripped off its surface in the
course of time. For the HVC, the situation is quite similar. Neutral gas is slowly
stripped off the outer wings of the cloud by momentum transfer between plasma
and neutral gas. The stripped gas atoms or molecules form a tail-like structure
with a higher temperature than the core of the cloud and a much smaller
density. The tail itself is laterally shielded by magnetic field lines draped
around the cloud forming a magneto-tail such that plasma and heat exchange
between tail and ambient medium are reduced. The simulated HVC
shows the same morphology and qualitatively similar density and temperature 
profiles as typical HVCs. Moreover,
the velocity pattern (Fig.~\ref{fig:vny55}) shows the same characteristics as the
observed HVC presented in Fig.~\ref{gauss125}: a fast moving cold core,
slower moving warm boundary layer and a slowly moving tail. \\
Finally,
Fig.~\ref{fig:nn55} shows the column density for the
neutral gas for a cut along the \(x\)-axis through the core of the cloud.
Here, we calculated the column density by rotating the cloud around the
\(y\)-axis and integrating along the line of sight lying in the
\(z\)-direction. This corresponds to measuring the column density by integrating
along the line of sight. 
Since we initially assumed invariance in the \(z\)-direction
this is of course not exact but reasonable for clouds which are not too small. 
The result
shows a smooth profile which is qualitatively very similar to column 
densities shown, for example, in Fig.~\ref{gauss125}. \\
All in all, the simulation shows the importance of the magnetic field for the
morphology of the HVC as well as for its stability. \\

\section{Discussion}

We presented the first plasma-neutral gas simulations of the interaction
of HVCs with the Galactic halo gas. Previous numerical studies
(Santilla\'an \cite{santi} and Quilis \cite{quilis}) focused on the dynamics in
external, Galactic gravitational fields. The questions of thermal and magnetic
insulation of the neutral gas clouds in the hot ambient medium as
well as their stability were not addressed. 
Our two-fluid approach provides rather detailed information
on the interaction region of the cold almost neutral HVC gas  and the hot
almost fully ionized ambient plasma. The Galactic magnetic field is locally
draped around the HVC, i.e., a magnetic barrier forms similar to magnetopauses
around comets (McComas \cite{comas}) and unmagnetized planets (Luhman \cite{luhma}). 
This barrier is of great importance
for the thermal insulation of the cold, fast traveling cloud 
from the hot surroundings. Furthermore, it prevents the plasma from diffusing
into the neutral gas cloud. We conclude that the formation of magnetic 
barriers contributes significantly to the dynamical stability of HVCs. 
Even for relatively weak initial magnetic fields in the Halo the magnetic
barrier sufficiently stabilizes the cloud.
In particular, unstable Kelvin-Helmholtz modes
are not excited due to the stabilizing effect of a magnetic field component
parallel to the flow. 
Thermal radiative losses that are not included in our contribution
may further enhance the stability of the
boundaries against Kelvin-Helmholtz modes (Vietri et al.~\cite{vie97}).
Within
the regions of discontinuity and in the adjacent inner layers the physics
becomes rather involved. In particular, some 
fraction of the stored magnetic field energy
can be converted into heat by magnetic reconnection and thus can be 
responsible for the observed x-ray emission (Zimmer et al.~\cite{zimmer}, 
Birk et al.~\cite{birk2}).
Moreover, one may note that the effect of ionization caused by the 
relative cloud halo movement (Konz et al.~\cite{konz}) may enhance the formation of
magnetic barriers by ionized boundary layers. \\ 
 A self-consistent three-dimensional simulation including all relevant
physical effects seems to be highly desirable and is a promising task for the 
future. 

Finally, we note that the interaction scenario studied in this paper may also 
be of relevance in the context of cloudlets in the magnetospheres of 
Active Galactic Nuclei as well as in cluster cooling flows.

\begin{acknowledgements}
      Part of this work was supported by the German
      \emph{Deut\-sche For\-schungs\-ge\-mein\-schaft, DFG\/} projects,
      LE 1039/5-1, LE 1039/6-1 and ME 745/19. \\
      We thank the referee for his helpful comments.

\end{acknowledgements}

\clearpage

\end{document}